\documentclass[twocolumn,showpacs,preprintnumbers,amsmath,amssymb,superscriptaddress]{revtex4-1}
\pdfoutput=1
\usepackage{latexsym,graphicx,rotating}
\usepackage{subfiles}
\usepackage{natbib}
\def\Nepg{{}^{22}\text{Ne}(p,\gamma)^{23}\text{Na}}
\def\Nang{{}^{23}\text{Na}(n,\gamma)^{24}\text{Na}}
\def\Neng{{}^{22}\text{Ne}(n,\gamma)^{23}\text{Ne}}
\def\Cpg{{}^{12}\text{C}(p,\gamma){}^{13}\text{N}}
\def\Alpg{{}^{27}\text{Al}(p,\gamma){}^{28}\text{Si}}

\begin{document}


\title{First inverse kinematics measurement of key resonances in the $\Nepg$ reaction at stellar temperatures} 

\author{A.~Lennarz}
\email[]{lennarz@triumf.ca}

\affiliation{TRIUMF, 4004 Wesbrook Mall, Vancouver, BC, Canada, V6T 2A3}
\author{M.~Williams}
\affiliation{TRIUMF, 4004 Wesbrook Mall, Vancouver, BC, Canada, V6T 2A3}
\affiliation{Department of Physics, University of York, Heslington, York, UK, YO10 5DD}
\author{A.~M.~Laird}
\affiliation{Department of Physics, University of York, Heslington, York, UK, YO10 5DD}
\affiliation{The NuGrid collaboration, http://www.nugridstars.org}
\author{U.~Battino}
\affiliation{The NuGrid collaboration, http://www.nugridstars.org}
\affiliation{University of Edinburgh, School of Physics and Astrophysics, Edinburgh EH9 3FD, UK}
\author{A.~A.~Chen}
\affiliation{Department of Physics and Astronomy, McMaster University, Hamilton, ON, Canada, L8S 4L8}
\author{D.~Connolly}
\altaffiliation[Present address: ]{Los Alamos National Laboratory, Los Alamos, New Mexico 87545, USA}
\affiliation{TRIUMF, 4004 Wesbrook Mall, Vancouver, BC, Canada, V6T 2A3}
\author{B.~Davids}
\affiliation{TRIUMF, 4004 Wesbrook Mall, Vancouver, BC, Canada, V6T 2A3}
\author{N.~Esker}
\altaffiliation[Present address: ]{San Jos\'e State University, 1 Washington Square, Duncan Hall 518 San Jos\'e, CA 95192-0101}
\affiliation{TRIUMF, 4004 Wesbrook Mall, Vancouver, BC, Canada, V6T 2A3}
\author{R.~Garg}
\altaffiliation[Present address: ]{University of Edinburgh, School of Physics and Astrophysics, Edinburgh EH9 3FD, UK}
\affiliation{Department of Physics, University of York, Heslington, York, UK, YO10 5DD}
\author{M.~Gay}
\affiliation{Columbia University, 116th St \& Broadway, New York, NY 10027, USA}
\author{U.~Greife}
\affiliation{Colorado School of Mines, Golden, CO, USA}
\author{U.~Hager}
\affiliation{The Joint Institute for Nuclear Astrophysics--Center for the Evolution of the Elements, Michigan State University, East Lansing, Michigan 48824, USA}
\author{D.~Hutcheon}
\affiliation{TRIUMF, 4004 Wesbrook Mall, Vancouver, BC, Canada, V6T 2A3}
\author{J.~Jos\'e}
\affiliation{Departament de F\'isica Universitat Polit\`ecnica de Catalunya \& Institut d'Estudis Espacials de Catalunya (IEEC), C. Eduard Maristany 10, E-08019 \& Ed. Nexus-201, C. Gran Capit\`a, 2-4, E-08034, Barcelona, Spain}
\author{M.~Lovely}
\affiliation{Colorado School of Mines, Golden, CO, USA}
\author{S.~Lyons}
\affiliation{The Joint Institute for Nuclear Astrophysics--Center for the Evolution of the Elements, Michigan State University, East Lansing, Michigan 48824, USA}
\affiliation{National Superconducting Cyclotron Laboratory, Michigan State University, East Lansing, MI  48824, USA}
\author{A.~Psaltis}
\affiliation{The NuGrid collaboration, http://www.nugridstars.org}
\affiliation{Department of Physics and Astronomy, McMaster University, Hamilton, ON, Canada, L8S 4L8}
\author{J.~E.~Riley}
\affiliation{Department of Physics, University of York, Heslington, York, UK, YO10 5DD}
\author{A.~Tattersall}
\affiliation{University of Edinburgh, School of Physics and Astrophysics, Edinburgh EH9 3FD, UK}
\affiliation{The NuGrid collaboration, http://www.nugridstars.org}
\author{C.~Ruiz}
\affiliation{TRIUMF, 4004 Wesbrook Mall, Vancouver, BC, Canada, V6T 2A3}

\date{\today}

\begin{abstract}
In this Letter we report on the first inverse kinematics measurement of key resonances in the $\Nepg$ reaction  which forms part of the NeNa cycle, and is relevant for ${}^{23}$Na synthesis in asymptotic giant branch (AGB) stars.  An anti-correlation in O and Na abundances is seen across all well-studied globular clusters (GC), however, reaction-rate uncertainties limit the precision as to which stellar evolution models can reproduce the observed isotopic abundance patterns. Given the importance of GC observations in testing stellar evolution models and their dependence on NeNa reaction rates, it is critical that the nuclear physics uncertainties on the origin of ${}^{23}$Na be addressed.
We present results of direct strengths measurements of four key resonances in $\Nepg$ at E$_{{\text c.m.}}$ = 149~keV, 181~keV, 248~keV and 458~keV. The strength of the important E$_{{\text c.m.}}$ = 458~keV reference resonance has been determined independently of other resonance strengths for the first time with an associated strength of $\omega\gamma$ = 0.439(22)~eV and with higher precision than previously reported. Our result deviates from the two most recently published results obtained from normal kinematics measurements performed by the LENA and LUNA collaborations but is in agreement with earlier measurements.
The impact of our rate on the Na-pocket formation in AGB stars and its relation to the O-Na anti-correlation was assessed via network calculations. Further, the effect on isotopic abundances in CO and ONe novae ejecta with respect to pre-solar grains was investigated.
\end{abstract}


\maketitle
Globular clusters (GCs) are dense aggregates of predominantly old stars found in the galactic halo. These objects have long fascinated astronomers for the unique insight they provide into the processes driving galaxy formation and chemical evolution. In particular, GCs are ideal test sites for answering open questions about the interplay between primordial and evolutionary chemical enrichment~\cite{gratton2004}. These objects have therefore warranted significant observational efforts and, through recent studies a complex picture of GCs abundance patterns has emerged, with strong evidence supporting multiple epochs of star formation~\cite{gratton2012}. Despite clear variability in observed abundances, some ubiquitous trends become apparent, such as the anti-correlation in oxygen and sodium abundances~\cite{Carretta-AA450-2006}. Currently stellar models are unable to reproduce many of the abundance patterns in GC stars along the red-giant branch (RGB), but absent in their field star counterparts. AGB stars undergoing Hot Bottom Burning (HBB) are currently the most favored astrophysical sites used to explain the O-Na anti-correlation~\cite{renzini2008,lee2010}.
HBB occurs during the quiescent phase between two thermal pulses (TP) when part of the H-shell is included in the envelope convection and the H-shell has enhanced access to fuel which is convectively mixed into its outer layers.
In TP-AGB stars, sodium is primarily synthesized  by  proton-capture  on ${}^{22}$Ne in the outer-most layer of the core-envelope transition zone, resulting in the formation of a so-called ${}^{23}$Na pocket on top of the ${}^{14}$N pocket~\cite{Cristallo-APJ696-2009, Lucatello-APJ-2011}. The ${}^{23}$Na pocket forms when ${}^{22}$Ne and ${}^{12}$C abundances are comparable, and the $\Nepg$ and $\Cpg$ reactions compete. In low-mass AGB stars, at solar metallicity, models predict the ${}^{23}$Na pocket to be the main sodium source, and the overproduction of Na to result from the ingestion of the ${}^{23}$Na pocket during the thermal dredge up~\cite{Cristallo-APJ696-2009}. The $\Nepg$ reaction further affects the ${}^{20}$Ne/${}^{22}$Ne, ${}^{21}$Ne/${}^{22}$Ne and ${}^{20}$Ne/${}^{21}$Ne abundance ratios of pre-solar grains found in meteorites. These grains are important signatures of nucleosynthesis in different stellar environments and mixing in stellar ejecta before the formation of our solar system.
The $\Nepg$ reaction is also influential in nova nucleosynthesis, where a sensitivity study by Iliadis~\cite{iliadis-AJS142-2002} showed that this reaction rate - varied within uncertainties - can affect the final abundances of $^{22}$Ne and $^{23}$Na by factors of $\sim$100 and $\sim$7, respectively.

In recent years the $\Nepg$ reaction has been targeted intensively by three facilities, all employing normal kinematics techniques~\cite{Depalo-PRC92-2015, Kelly-PRC95-2017, Ferraro-PRL121-2018, Cavanna-PRL120E-2018}. The low-energy regime was investigated by the LUNA and LENA collaborations, since the rate is dominated by narrow low-energy resonances. With the exception of the low-energy resonance strength measurements by LUNA~\cite{Ferraro-PRL121-2018, Cavanna-PRL120E-2018} with E$_{{\text c.m.}} \leq$ 248~keV, all previously reported strengths were either measured relative to reference resonances at E$_{{\text c.m.}}$ = 458~keV or 1222~keV or depended on these resonances to determine target stoichiometries. 
The 458~keV resonance strength directly influences the strengths of the low-energy resonances reported by LENA~\cite{Kelly-PRC95-2017}, and was used as reference for target stoichiometries in ${}^{22}$Ne+$\alpha$ studies~\cite{Giesen-NPA561-1993} and normal kinematics measurements of the $\Nepg$ reaction~\cite{Depalo-PRC92-2015}. 
Moreover, the 458~keV resonance is particularly relevant for reaction-rate compilations conducted by Iliadis {\it et al.}~\cite{Iliadis-NPA841-2010}, for which all other measured resonance strengths were normalized to the 458~keV strength value of $\omega\gamma$ = 0.524(51)~eV~\cite{Longland-PRC81-2010}. The latter was determined relative to the E$_{{\text p}}$ = 405.5(3)~keV ($\omega\gamma$ = (8.63(52)$\times$10$^{-3}$)~eV~\cite{Powell-NPA644-1998}) resonance strength in the $\Alpg$ reaction, and depends on the background contribution of the E$_{{\text p}}$ = 326~keV and 447~keV resonances in the same reaction. We note that there is a more recent result for the E$_{{\text p}}$ = 405.5(3)~keV resonance of $\omega\gamma$ = 1.04(5)$\times$10$^{-2}$~eV~\cite{Harissopulos-EPJA9-2000}. Using this value for a linear re-normalization would reduce the 458~keV strength reported by Longland~{\it et al.} by $\sim$17$\%$ to $\omega\gamma$ = 0.435(42)~eV. Further, the strengths of the resonances affecting the background in that measurement have also been normalized to the 405.5(3)~keV resonance. Though the 458~keV resonance strength has been investigated numerous times~\cite{Longland-PRC81-2010, Depalo-PRC92-2015, Kelly-PRC92-2015}, our measurement reveals that the situation for this resonance is still not resolved. In fact, the strength of this resonance has never been measured independently of other resonances. However, this work puts forward a direct, reference-independent measurement which is largely independent of knowledge of the relevant branching ratios (BRs).

In this Letter we report on the first inverse kinematics measurement of the $\Nepg$ reaction, which comprises the strength determination of the to-date largest set of resonances for this reaction measured within one experiment, covering an energy range from E$_{{\text c.m.}}$ = 149~keV to 1.222~MeV.

The measurement was performed using the DRAGON (Detector of Recoil and Gammas Of Nuclear reactions) recoil separator~\cite{Hutcheon-NIMPRSA498-2003} located at the ISAC beam facility at TRIUMF, Vancouver, Canada. DRAGON is designed to conduct studies of radiative capture reactions in inverse kinematics and consists of three main sections: (1) a windowless, differentially pumped, recirculated gas target surrounded by a high-efficiency $\gamma$-detector array consisting of 30 BGO detectors; (2) a high-suppression electromagnetic mass separator with two stages of charge and mass selection; (3) a variable heavy ion detection system in combination with two micro-channel plate (MCP) based timing detectors for time-of-flight (TOF) measurements. The recoil-detection system consisted of a double-sided silicon strip detector (DSSSD)~\cite{Hutcheon-NIMPRSA498-2003,Vockenhuber-NIMPRB266-2008}.

\begin{figure}[ttt]
\includegraphics[width=0.45\textwidth]{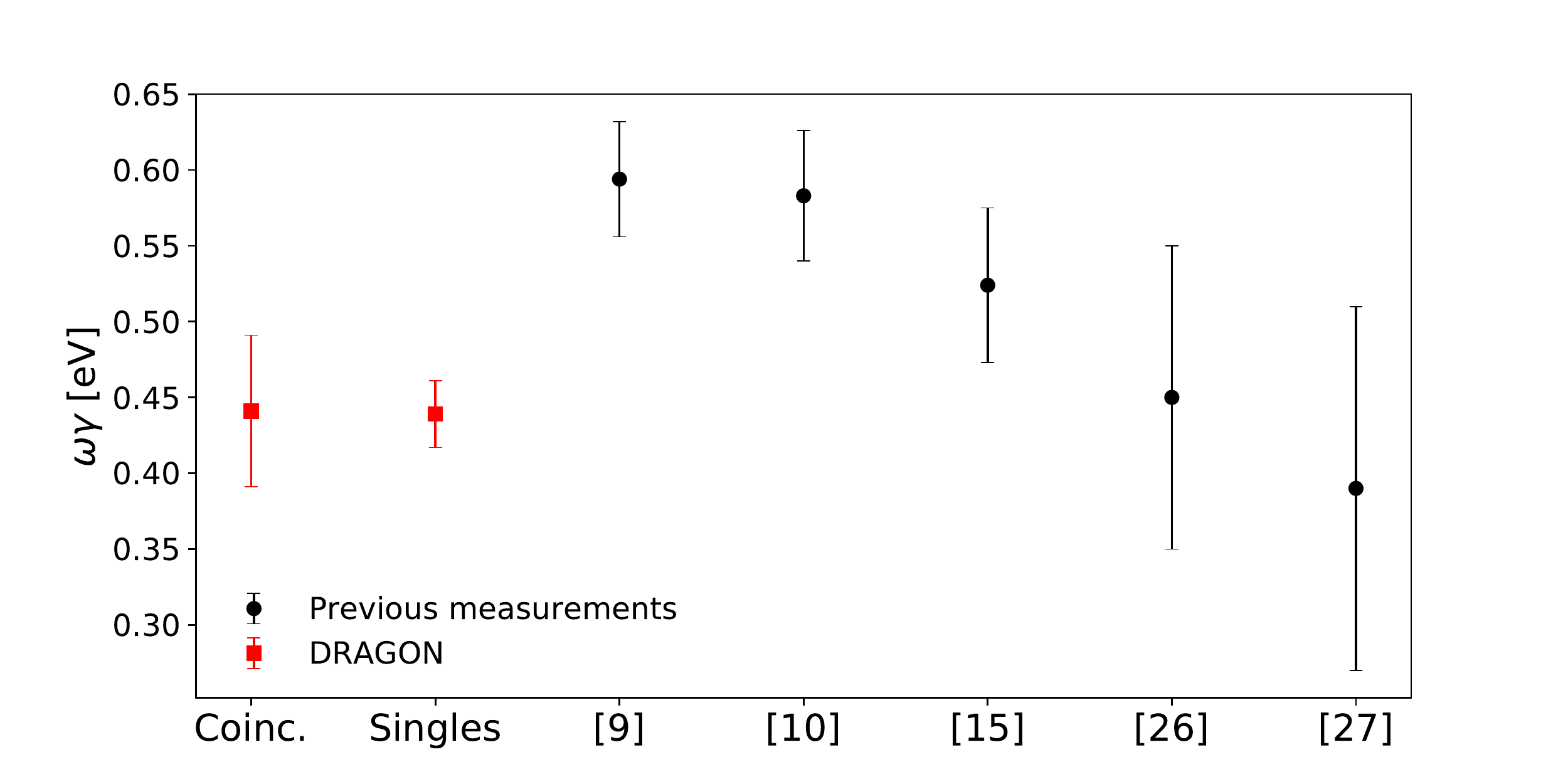}
\caption{Previous 458~keV strength values (black circles) in relation to the DRAGON results (red squares) obtained from singles and coincidence analysis.\label{fig:results}}
\end{figure}

A high intensity ($\sim$2 $\times$ 10$^{12}$~ions/sec) isotopically pure ${}^{22}$Ne$^{4+}$ beam was delivered to the hydrogen-filled gas target. $^{23}$Na recoils were transmitted through the separator and detected in the DSSSD. To contain the entire yield profile of the resonances within the target, an average H$_{2}$ gas pressure of 5~Torr was used, corresponding to a target thickness of $\sim$3.9$\times$10$^{18}$ hydrogen atoms/cm$^{2}$. The maximum charge state at each energy was selected by transmitting the beam through the magnetic dipoles. Charge-state distributions for ${}^{23}$Na ions in hydrogen gas at the recoil energies were measured to eliminate systematic uncertainties associated with semi-empirical calculations. Two silicon surface barrier detectors positioned at 30$^{\circ}$ and 57$^{\circ}$ relative to the beam axis inside the target detected elastically scattered protons for a relative measure of the beam intensity. The elastic scattering rate was normalized to automated hourly Faraday Cup readings.
Prior to each yield measurement, the energy loss across the target was determined by measuring the incoming and outgoing beam energy via the magnetic field of the first magnetic dipole, which centered the beam on-axis on a pair of current sensitive slits. The incoming beam-energy spread was $\sim$0.1$\%$ FWHM~\cite{Hutcheon-NIMPRA-2012}. Stopping powers ($\epsilon$) were calculated based on the measured energy loss, the gas density derived from continuously recorded pressure and temperature, and the effective target length~\cite{Hutcheon-NIMPRSA498-2003}. This reduces uncertainties induced by the commonly used software packages SRIM~\cite{Ziegler-NIMRSB268-2010} and LISE~\cite{Kuchera-JPCS664-2015}. Resonance energies were determined via the position sensitive BGO array by relating the centroid of the distribution ($\gamma$ yield vs target position) to the incoming and outgoing beam energy~\cite{Hutcheon-NIMPRA-2012}.
\begin{figure}[ttt]
 \includegraphics[width=0.4\textwidth]{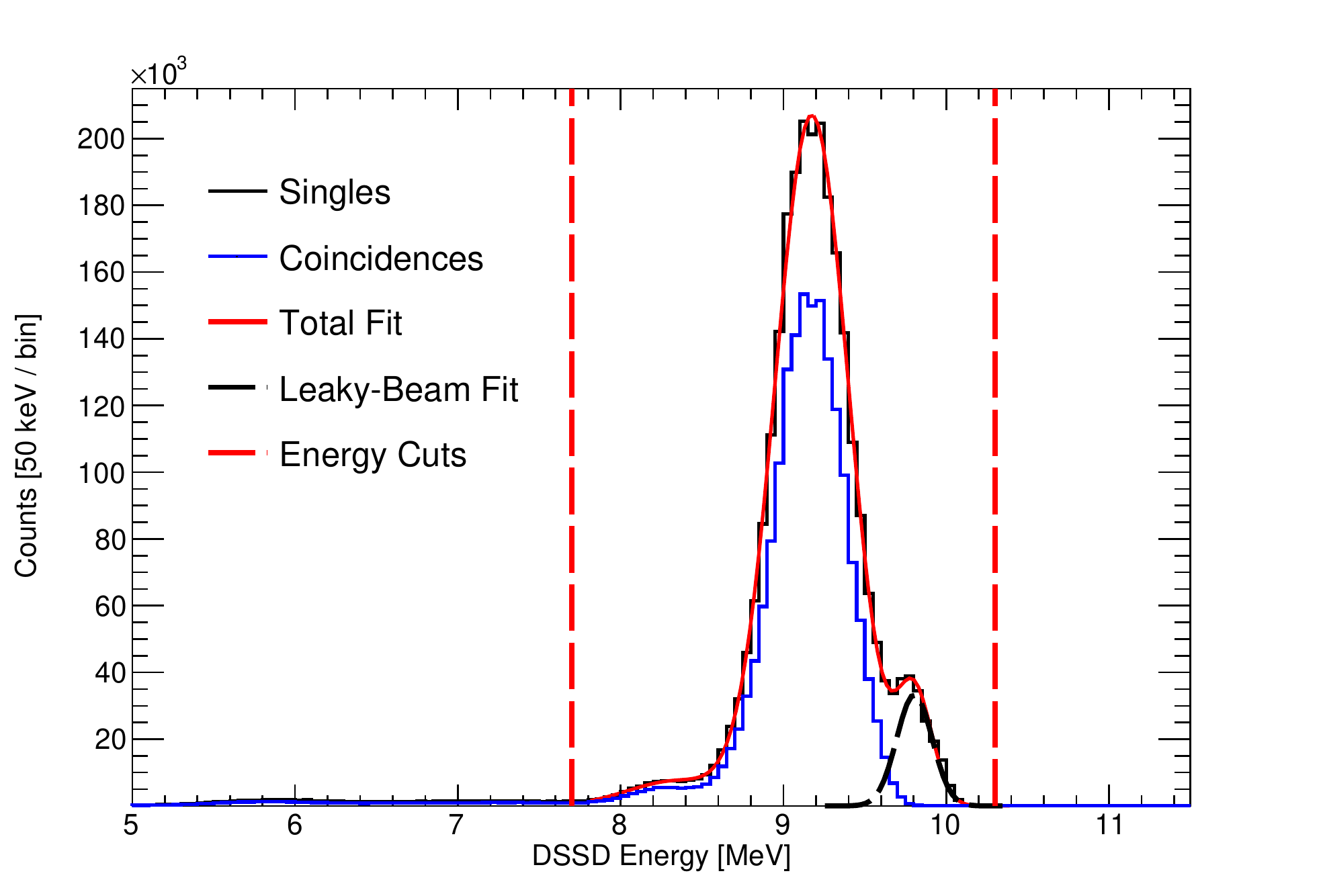}
\caption{Singles (black histogram) and coincidence (blue histogram) DSSSD energy spectra of the 458~keV yield measurement. In red, the triple Gaussian fit of the singles spectrum is shown. The black dashed line denotes the unreacted beam component (not present in coincidence measurement) of the fit and the red dashed vertical lines indicate the recoil gate.}
\label{fig:DSSSD}
 \end{figure}
For improved background suppression, the resonance strengths were extracted in a coincidence analysis, where the \texttt{GEANT3}~\cite{Brun:1987ma} simulation used to determine the BGO detection efficiency relies on literature BRs.
For the 458~keV yield measurement the DSSSD energy spectrum was fitted with a double Gaussian function to set appropriate energy cuts for the ''golden'' recoil gate at $\pm$3.5$\sigma$ relative to the peak centroids, and to account for the satellite peak at the low energy side of the main recoil peak (Fig.~\ref{fig:DSSSD}). The satellite peak results from the additional energy loss of ions passing the 3$\%$ aluminum DSSSD grid~\cite{Wrede-NIMB204-2003}. Including the satellite peak and accounting for inter-strip events results in a DSSSD efficiency of (96.15 $\pm$ 0.1$_{stat.}$ $\pm$ 0.43$_{sys.}$)$\%$~\cite{Wrede-NIMB204-2003}.
The established DSSSD and BGO energy gates were then placed on the separator TOF spectrum to extract the number of recoils. The background within the recoil region was estimated by sampling the time-random background and calculating an average expectation value over the width of the signal region. A poissonian background model was chosen as the probability to count a random coincidence in the separator TOF spectrum follows Poisson statistics.
High statistics and a clear separation of unreacted beam and recoils (Fig.~\ref{fig:DSSSD}) also allowed for a singles analysis of the 458~keV resonance to eliminate uncertainties introduced by the dependence of the coincidence analysis on BRs and BGO detection efficiency. Using the fit parameters of the coincidence spectrum as guide for the singles analysis, a triple Gaussian function was applied to the DSSSD energy spectrum, and the integral of the main recoil peak and satellite peak comprises the number of recoil events.
Figure~\ref{fig:results} presents the 458~keV resonance-strength values based on coincidence and singles analysis, which are mutually consistent, relative to previous measurements.

Our result for the 458~keV resonance strength of $\omega\gamma_{coinc}$ = 0.441(50)~eV ($\omega\gamma_{singles}$ = 0.439(22)~eV) is lower and not in agreement within errors with the two latest results~\cite{Depalo-PRC92-2015, Kelly-PRC92-2015}. However, it is in agreement with three previous values~\cite{Longland-PRC81-2010, Endt-NPA521-1990, Meyer-NPA205-1973}. The result from Meyer {\it et al.}~\cite{Meyer-NPA205-1973} was normalized to the E$_{{\text c.m.}}$ = 612~keV resonance strength, and the Endt~{\it et al.} value is based on Ref.~\cite{Meyer-NPA205-1973}, however, normalized the E$_{{\text c.m.}}$ = 1.222~MeV resonance strength from Ref.~\cite{Keinonen-PRC15-1977}. The sensitivity of former studies to reference resonances underlines the necessity of reference-independent measurements as well as more precise measurements of reference-resonance strengths.

\begin{figure*}[ttt]
\minipage{0.3\textwidth}
  \includegraphics[width=1.1\textwidth]{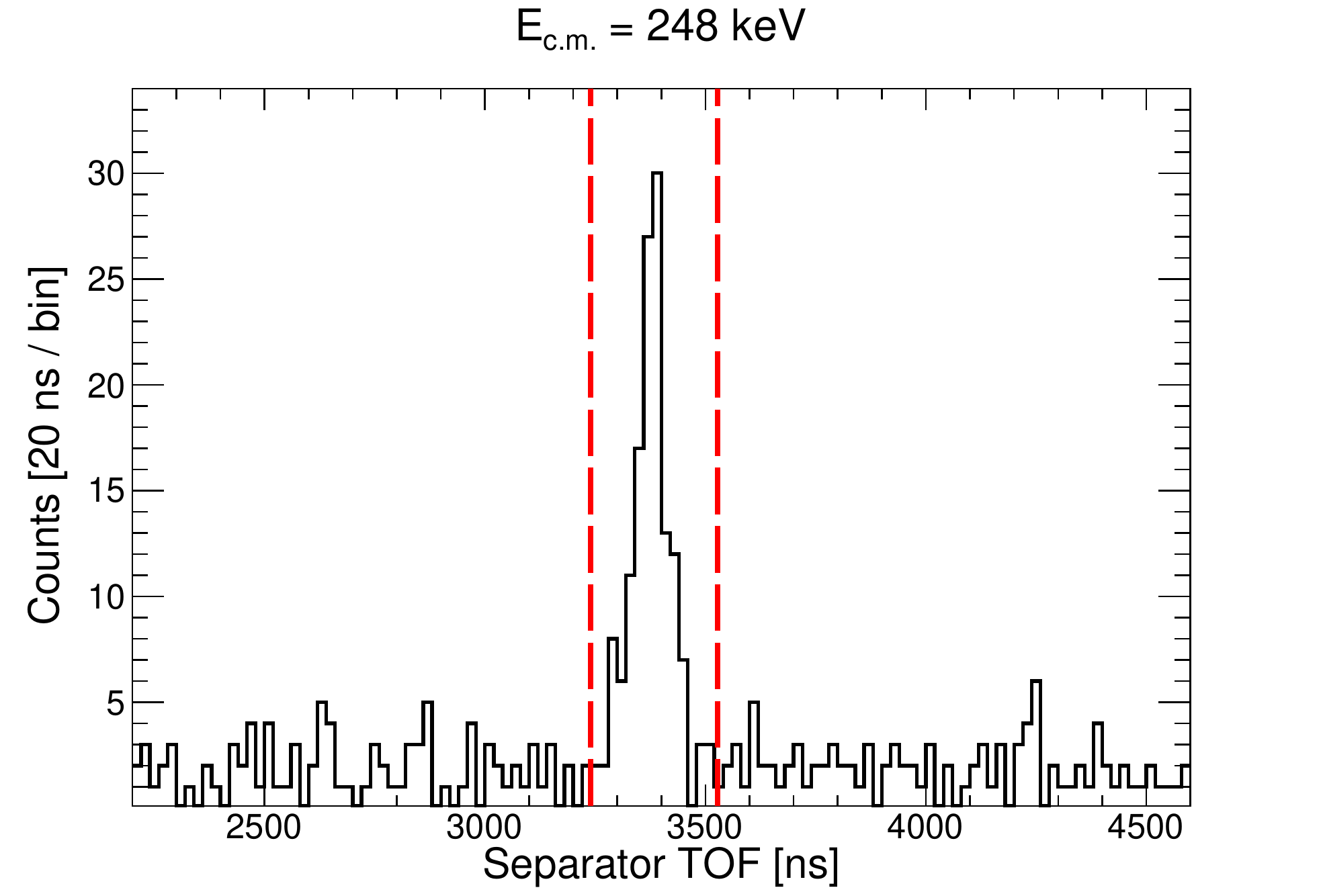}
\endminipage\hfill
\minipage{0.3\textwidth}
  \includegraphics[width=1.1\textwidth]{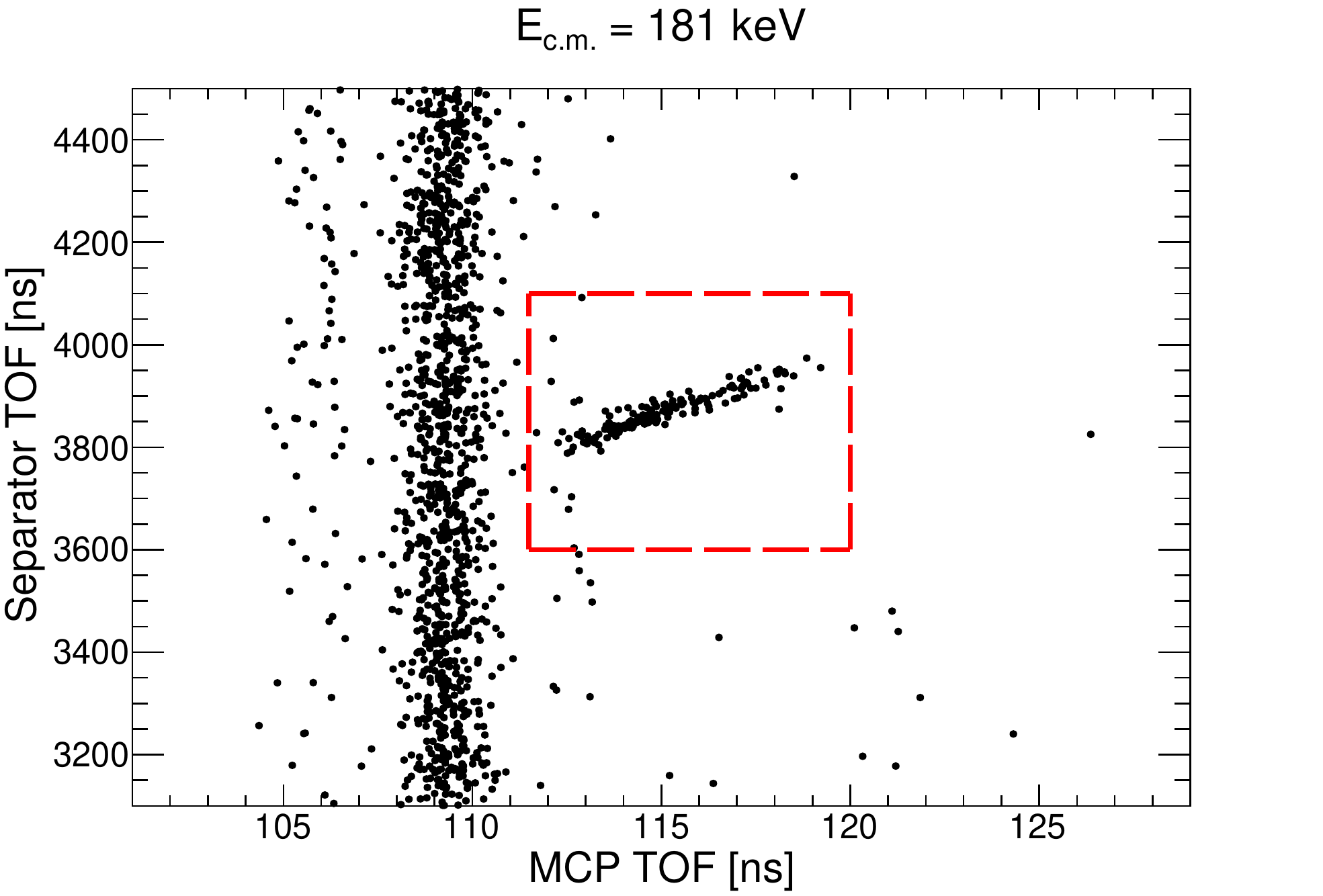}
\endminipage\hfill
\minipage{0.3\textwidth}
  \includegraphics[width=1.1\textwidth]{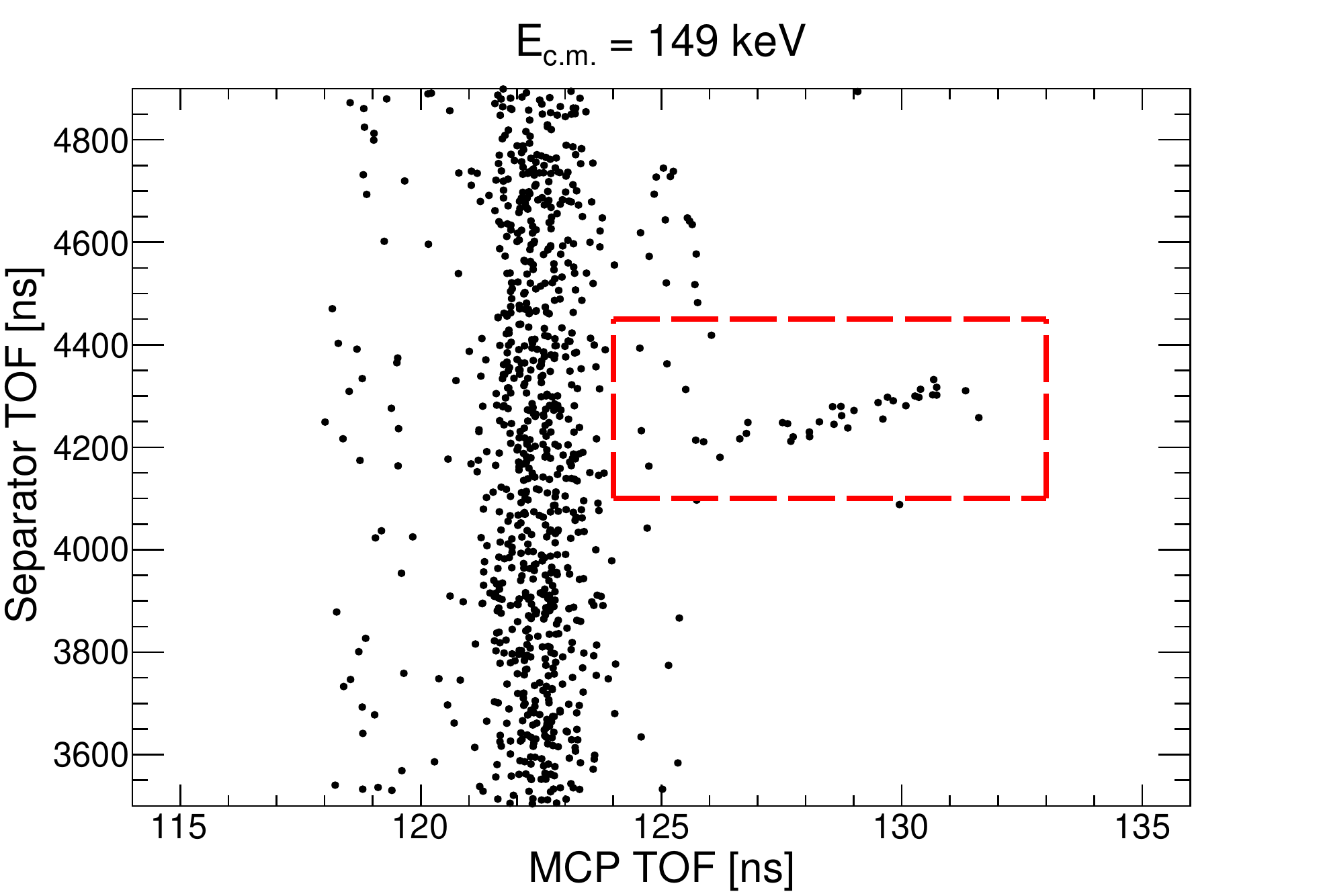}
\endminipage
\caption{Separator TOF spectrum for the E$_{{\text c.m.}} =$ 248~keV resonance, and separator vs MCP TOF spectra for the E$_{{\text c.m.}} = $181~keV and 149 keV yield measurements. The red dashed lines represent the recoil timing gates. Each spectrum is gated on the recoil peak in the DSSSD energy spectrum and a minimum BGO energy threshold of $E_{\gamma} >$ 2.2, 2.0, and 2.5 MeV, respectively.}
\label{fig:xtof}
\end{figure*}
To determine the 149~keV, 181~keV and 248~keV resonance strengths, conservative recoil gates for DSSSD and BGO energy were placed on the separator TOF vs MCP TOF spectrum or separator TOF spectrum (Fig.~\ref{fig:xtof}). The 248~keV yield measurement does not have an associated separator vs MCP TOF spectrum since the MCP detection efficiency was too low to give enough statistics; this issue was later resolved for the lower energy measurements.

For the analysis of the 149~keV and 181~keV yield measurements the branching ratios for the E$_{x}$ = 8943(3)~keV and 8972(3)~keV levels given in Ref.~\cite{Kelly-PRC95-2017} were used for the \texttt{GEANT3} simulation. The BRs from Ref.~\cite{Kelly-PRC95-2017} were chosen over those reported in Ref.~\cite{Depalo-PRC94-2016} as the analysis in Ref.~\cite{Kelly-PRC95-2017} did not require additional background subtraction or corrections for coincidence-summing effects, and accounted for escape peaks and Compton continuum.

For the 149~keV resonance we report a strength of $\omega\gamma$(149) = (1.67 $\pm$ 0.028 (sys) $^{+0.039}_{-0.028}$ (stat))$\times10^{-7}$~eV, which is lower but in agreement with all previously published values. Our 181~keV resonance strength of $\omega\gamma$(181) = (2.17$^{+0.32}_{-0.31}$ (sys) $^{+0.2}_{-0.17}$ (stat))$\times10^{-6}$~eV is in good agreement with the LUNA HPGe result~\cite{Cavanna-PRL120E-2018} and lower but also in agreement with the LENA result. However, our result is 20$\%$ lower than the LUNA BGO measurement~\cite{Ferraro-PRL121-2018} (compare Tab.~\ref{tab:wg}).
Regarding the 248~keV resonance we report a strength of $\omega\gamma$ = 8.5(1.4)$\times$10$^{-6}$~eV.
The dominant contributions to the systematic uncertainty result from uncertainties on coincidence efficiency (10$\%$), stopping power (4.3 - 5.9$\%$), charge-state fraction (1.8$\%$(181~keV) - 2.4$\%$(149~keV)), MCP efficiency (5$\%$) and beam normalization (1.1 - 4.9$\%$).   
To clarify that there is no trend in systematically lower strengths values relative to the LUNA results, we note that the DRAGON results for higher energy resonances are either in agreement with previous results or slightly higher, and will be published elsewhere~\cite{Williams-PRC-2019}.
 \begin{table}
 \caption{Overview of resonance strengths. (S) marks results from a singles analysis. \label{tab:wg}}
 \begin{ruledtabular}
 \begin{tabular}{l l l l }
 E$_{\text{c.m.}}$[keV] & \multicolumn{2}{c}{$\omega\gamma$ [eV]}\\
 & Lit. &  \multicolumn{2}{l}{This work} \\
 458 & 0.583(43)~\cite{Kelly-PRC92-2015} &0.441(50) & 0.439(22) (S) \\
     & 0.605(61)~\cite{Depalo-PRC92-2015} & &   \\
 248 & 8.2(7)$\times$10$^{-6}$~\cite{Cavanna-PRL120E-2018} & \multicolumn{2}{c}{8.5(1.4)$\times$10$^{-6}$}   \\
   & 9.7(7)$\times$10$^{-6}$~\cite{Ferraro-PRL121-2018} &  &  \\
 181 & 2.2(2)$\times$10$^{-6}$~\cite{Cavanna-PRL120E-2018} & & \\
 & 2.7(2)$\times$10$^{-6}$~\cite{Ferraro-PRL121-2018} & \multicolumn{2}{c}{2.17$^{+0.37}_{-0.35} \times$10$^{-6}$}  \\
  & 2.32(32)$\times$10$^{-6}$~\cite{Kelly-PRC95-2017} & \multicolumn{2}{c}{}  \\
 149 & 1.8(2)$\times$10$^{-7}$~\cite{Cavanna-PRL120E-2018} & & \\
         & 2.2(2)$\times$10$^{-7}$~\cite{Ferraro-PRL121-2018} & \multicolumn{2}{c}{1.67$^{+0.48}_{-0.40}\times$10$^{-7}$}   \\
          & 2.03(40)$\times$10$^{-7}$~\cite{Kelly-PRC95-2017} & \multicolumn{2}{c}{}  \\
 &  \multicolumn{3}{c}{Ref.~\cite{Kelly-PRC95-2017} renormalized to this work}\\    
181 & \multicolumn{2}{r}{1.75(29)$\times$10$^{-6}$} \\
149 & \multicolumn{2}{r}{1.53(33)$\times$10$^{-7}$}\\
 \end{tabular}
 \end{ruledtabular}
 \end{table}
In view of the significant deviation of the DRAGON $\omega\gamma$(458~keV) result from the value used to normalize the strengths of the low-energy resonances in the TUNL measurement~\cite{Kelly-PRC95-2017}, one has to carefully review the latter. In fact, re-normalizing the LENA 149~keV strength to our $\omega\gamma$(458~keV) result, brings it into better agreement with the DRAGON measurement, and a re-normalized LENA value for the 181~keV resonance is compatible with the DRAGON and LUNA HPGe results.

\begin{figure}[ttt]
\includegraphics[width=0.37\textwidth]{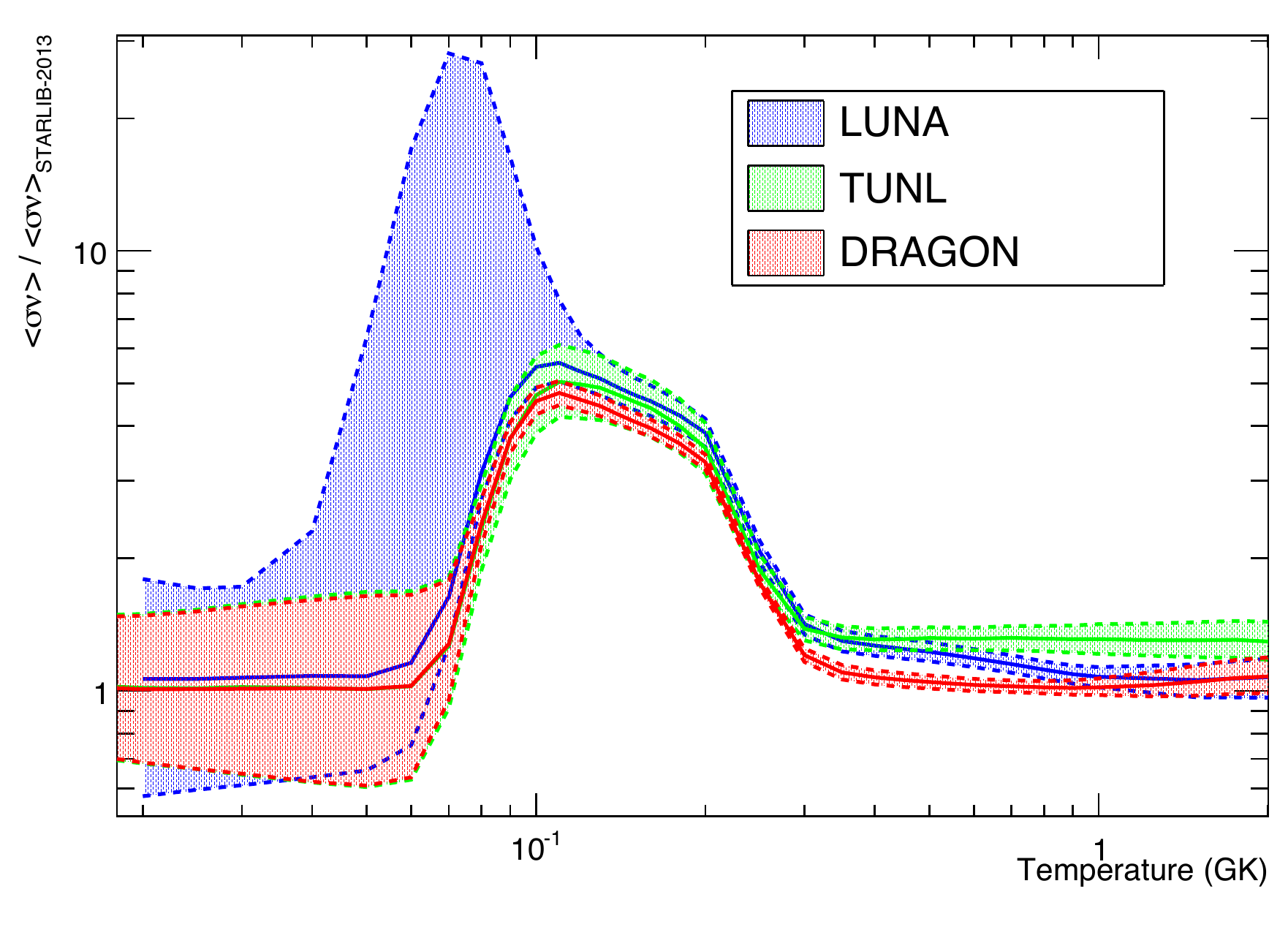}
\caption{The $\Nepg$ reaction rate normalized to the \texttt{STARLIB2013} rate~\cite{Sallaska-TAJSS207-2013}. Shaded areas bound the $1-\sigma$ upper and lower limits of each calculated rate. The DRAGON rate was calculated using the same \texttt{RateMC} code~\cite{Iliadis-NPA841-2010} used for the LENA rate.}
\label{fig:rate}
\end{figure}

Figure~\ref{fig:rate} displays an overlay of the rates determined from this work and those of LUNA and LENA, normalized to the \texttt{STARLIB2013} rate~\cite{Sallaska-TAJSS207-2013}. The dramatic enhancement of the LUNA rate is mainly due to the inclusion of the E$_{{\text c.m.}}$ = 100~keV resonance, for which only an upper limit has been reported~\cite{Cavanna-PRL115-2015, Ferraro-PRL121-2018}. Our rate maps closely with the LENA rate, with a slight reduction due to our reduced 149~keV and 181~keV strengths.

\begin{figure}[ttt]
\includegraphics[width=0.37\textwidth]{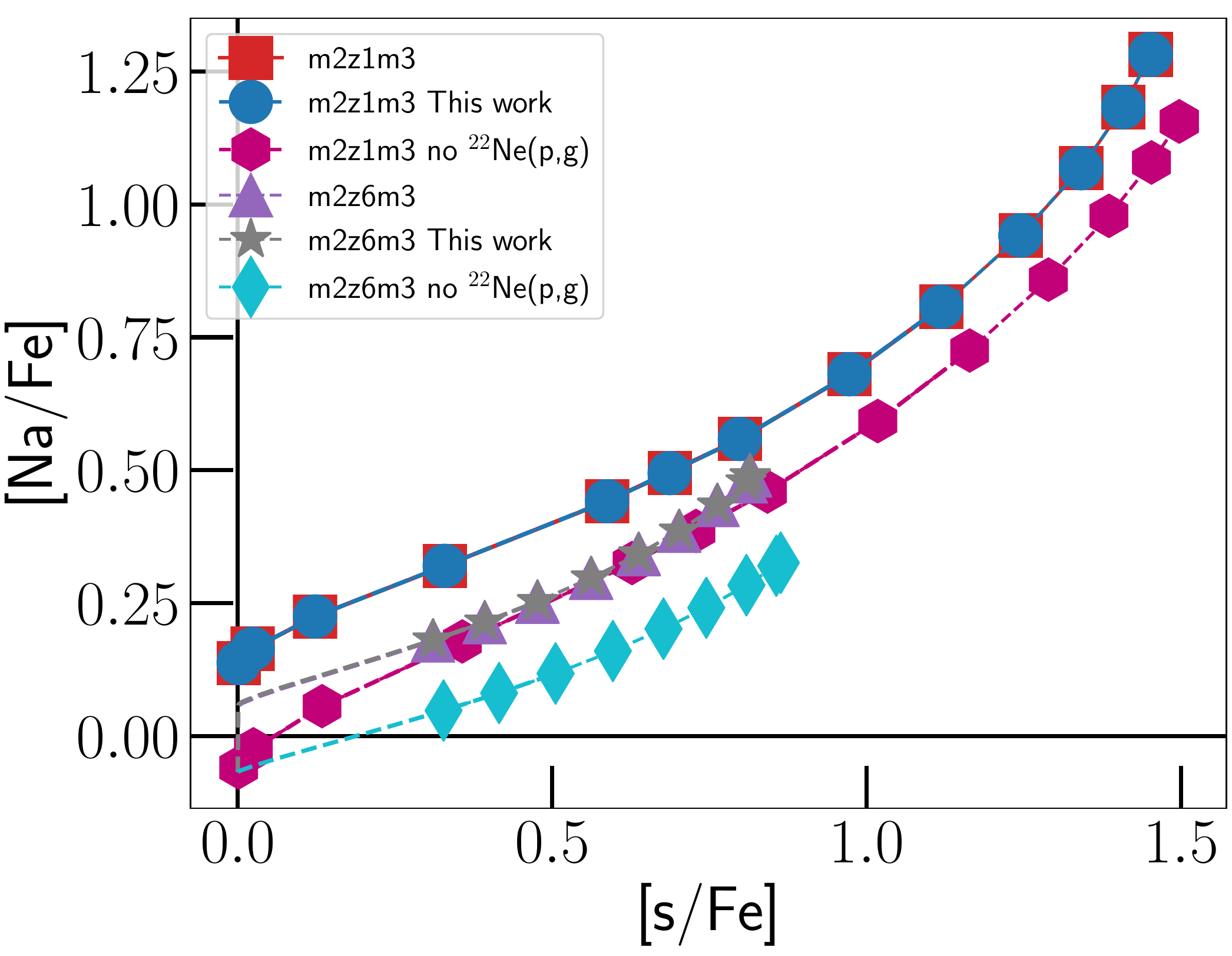}
\caption{Predicted surface [Na/Fe] abundance ratio as a function of {\it s} process element abundances [s/Fe] for a 2M$_{\odot}$ AGB star using the rate from this work.}
\label{fig:abundance}
\end{figure}

The effect of the DRAGON rate compared to the Iliadis 2010 rate~\cite{Iliadis-NPA841-2010} on the sodium and neon abundances in neon-oxygen (ONe) novae with underlying white-dwarf (WD) masses of 1.15~M$_{\odot}$ and 1.25~M$_{\odot}$, as well as carbon-oxygen (CO) novae (1.15~M$_{\odot}$ and 1.00~M$_{\odot}$) was investigated using hydro-dynamical nova models~\cite{Jose-APJ494-1998, Jose-SE-2016}. Changes of more than 10$\%$ in the isotopic abundances within the Ne-Al region (${}^{20,21,22}$Ne, ${}^{22,23}$Na, ${}^{25,26}$Mg, ${}^{26,27}$Al) in 1.15~M$_{\odot}$ CO novae, and a factor of 2 enhancement in ${}^{23}$Na abundance are observed for both CO nova models. For ONe novae, a reduction of the ${}^{22}$Ne content by a factor of 2 is observed for both WD mass models. Further, the ${}^{24}$Mg abundance is enhanced by $\sim$15$\%$ in the 1.25~M$_{\odot}$ model, whereas only slight differences are seen for the remaining isotopes considered in both models. 
Regarding CO novae, the new rate underlines the differences in the ${}^{25}$Mg/${}^{26}$Mg and ${}^{26}$Mg/${}^{25}$Mg ratios between the 1.0 and 1.15~M$_{\odot}$ models. Our rate leads to an increase of 24$\%$ in the isotopic ratio of ${}^{25}$Mg/${}^{24}$Mg, and to a decrease of 13$\%$ in the ${}^{26}$Mg/${}^{25}$Mg ratio for the 1.15~M$_{\odot}$ model relative to the \texttt{STARLIB2013} rate. This can be explained by the sensitivity of Mg synthesis to the peak temperature~\cite{Jose-APJ414-2004}. Due to the larger rate the mass flow is pushed up to Mg synthesis temperatures. As a result of this correlation these ratios become relevant in the identification of pre-solar grains which have a putative CO novae origin, as they function as probe for the peak temperature reached in the outburst, and the underlying WD mass. In a sensitivity study~\cite{iliadis-AJS142-2002}, the final abundances of ${}^{24,25}$Mg for 1.15~M$_{\odot}$ CO novae varied by up to a factor of 5, when varying the $\Nepg$ rate within its uncertainties, whereas the new rate strongly limits the reaction rate uncertainty in the temperature range of interest (T$_{peak}$ = 170~MK). Varying the new rate within its limits only changes the Mg isotope mass fractions by up to 7$\%$. For ONe novae, the cycling back to ${}^{20}$Ne is irrelevant for both the 1.15 and 1.25~M$_{\odot}$ model, as ${}^{20}$Ne is sufficiently available. This is reflected in the same ${}^{20,21}$Ne final yield, independent of the model. Though differences in the ${}^{22}$Ne abundance are found, abundances of ${}^{23}$Na, ${}^{24}$Mg or higher mass isotopes remain unaffected. Instead, the observed difference in  ${}^{22}$Ne abundances may be relevant for studies of pre-solar grains, which are identified by noble gas ratios.
The NuGrid multi-zone post-processing code MPPNP~\cite{Ritter-MNRAS-2018} was used to implement our rate in nucleosynthesis network calculations, and to model the [Na/Fe] abundance ratio on the AGB star surface at the end of the evolution of stable isotopes for various masses and metallicities.
A 5~M$_{\odot}$ model with metallicity z = 0.006 was utilized to study the impact of our rate on HBB in TP-AGB stars, using the \texttt{STARLIB2013} rate as reference. We observe a close mapping of [Na/Fe] as a function of [s/Fe] for the two rates, confirming the robustness of the \texttt{STARLIB2013} rate. The effect of the $\Nepg$ and $\Neng$ rates on the sodium abundance was studied. Without the (p,$\gamma$) channel, the abundance drops to almost zero, confirming the $\Nepg$ reaction as main production channel of sodium in massive AGB stars.
Further, the effect on the ${}^{23}$Na-pocket in low-mass AGB stars for a 2M$_{\odot}$ model (at Z = 0.001 and Z = 0.006) formed with the DRAGON rate relative to the \texttt{STARLIB2013} rate was investigated by evaluating the abundance profile of ${}^{23}$Na when the pocket is fully formed (Fig.~\ref{fig:abundance}).
Switching off the $\Nepg$ reaction results in a significant abundance reduction. However, in contrast to the 5~M$_{\odot}$ model, the abundance stays relatively high due to the second production channel ${}^{22}$Ne(n,$\gamma$)${}^{23}$Ne($\beta^{-}$)${}^{23}$Na. 
At T = 100~MK, we find a factor of 4 enhanced rate relative to the \texttt{STARLIB2013} rate. However, the differences between the results obtained with the $\Nepg$ \texttt{STARLIB2013} and DRAGON rate are minor, showing that the abundance does not directly correlate with rate variations.
This is not in agreement with the factor of 3 enhancement in ${}^{23}$Na production stated by Slemer~{\it et.~al.}~\cite{Slemer-MNRAS465-2017} based on the LUNA rate, which includes the tentative 68~keV and 100~keV resonances. Even though Slemer~{\it et.~al.} use a code that couples mixing and burning during HBB, and adopt a similar list of isotopes as NuGrid, neutron captures are not included. Thus, the important ${}^{23}$Na destruction channel $\Nang$ stated in Ref.~\cite{Cristallo-MSAI77-2006} is not considered.

In summary, key resonances in the $\Nepg$ reaction have been investigated in inverse kinematics for the first time using the DRAGON recoil separator. The strength of the important reference resonance at E$_{{\text c.m.}}$= 458~keV has been determined with higher precision via a direct measurement, and does not agree within errors with the two most recent normal kinematics results. Our result affects resonance strengths that have been determined relative to the strength of this resonance, as well as neon-target stoichiometries determined based on its strength. A new reaction rate was calculated based on the DRAGON measurement, which confirms the accuracy of the current ${}^{23}$Na production results in AGB stars in relation to the behavior of the $\Nepg$ reaction and underlines the importance of this reaction for the sodium production in AGB stars. Further work is needed to reassess the sensitivity of Mg isotopic ratios in CO novae to rate variations of isotopes in the Ne-Al region to use said ratios as a probe of the underlying WD peak temperatures.

\begin{acknowledgments}
The authors thank the ISAC operations and technical staff at TRIUMF. TRIUMF’s core operations are supported via a contribution from the federal government through the National Research Council Canada, and the Government of British Columbia provides building capital funds. DRAGON is supported by funds from the National Sciences and Engineering Research Council of Canada. The authors acknowledge support from the ''ChETEC'' COST Action (CA16117), supported by COST 116 (European Cooperation in Science and Technology). MW, AML, JR were supported by the UK Science and Technology Facilities Council (STFC). UB acknowledges support from the European Research Council ERC-2015-STG Nr. 677497. J. Jos\'{e} acknowledges support from the Spanish MINECO grant AYA2017-86274-P, the EU FEDER funds and the AGAUR/Generalitat de Catalunya grant SGR-661/2017. Authors from the Colorado Scool of Mines acknowledge funding via the U.S. Department of Energy grant DE-FG02-93ER40789. The authors also thank R. Longland for his support in calculating the thermonuclear reaction rate presented in this work.
\end{acknowledgments}

\bibliography{references22Ne.bib}

\end{document}